\documentstyle[preprint,aps]{revtex}
\tighten
\draft
\widetext
\input epsf
\preprint{CLNS 97/1491, HUTP-97/A026, NUB 3161}
\bigskip
\bigskip
\begin{document}
\title{4D Chiral $N=1$ Type I Vacua with and without $D5$-branes}
\medskip
\author{Zurab Kakushadze$^{1,2}$\footnote{E-mail: 
zurab@string.harvard.edu} and 
Gary Shiu$^3$\footnote{E-mail: shiu@mail.lns.cornell.edu}}

\bigskip
\address{$^1$Lyman Laboratory of Physics, Harvard University, Cambridge, 
MA 02138\\
$^2$Department of Physics, Northeastern University, Boston, MA 02115\\
$^3$Newman Laboratory of Nuclear Studies, Cornell University,
Ithaca, NY 14853-5001}
\date{June 8, 1997}
\bigskip
\medskip
\maketitle

\begin{abstract}
{}In this paper we consider compactifications of type I strings on Abelian 
orbifolds. We discuss the tadpole cancellation conditions for the general 
case with $D9$-branes only. Such compactifications have (perturbative) 
heterotic duals which are also realized as orbifolds (with non-standard 
embedding of the gauge connection). The latter have extra twisted states 
that become massive once orbifold singularities are blown-up. This is due to 
the presence of {\em perturbative} heterotic superpotential with couplings 
between the extra twisted states, the orbifold blow-up modes, and (sometimes) 
untwisted matter fields. Anomalous $U(1)$ (generically present in such 
models) also plays an important role in type I-heterotic (tree-level) duality 
matching. We illustrate these issues on a particular example of 
${\bf Z}_3\otimes {\bf Z}_3$ orbifold type I model (and its heterotic dual). 
The model has $N=1$ supersymmetry, $U(4)^3\otimes SO(8)$ gauge group, and 
chiral matter. We also consider compactifications of type I strings on 
Abelian orbifolds with both $D9$- and $D5$-branes. We discuss tadpole 
cancellation conditions for a certain class of such models. We illustrate 
the model building by considering a particular example of type I theory 
compactified on 
${\bf Z}_6$ orbifold. The model has $N=1$ supersymmetry, 
$[U(6)\otimes U(6)\otimes U(4)]^2$ gauge group, and chiral matter. 
This would correspond to a non-perturbative chiral vacuum from the
heterotic point of view.

\end{abstract}

\section{Introduction}

{}Recently it has become clear that type I-heterotic 
duality \cite{typeI-het-10} can be extended to $N=1$ cases in four 
dimensions \cite{z3,z7}. In particular, two cases with $D9$-branes 
(but no $D5$-branes) have been worked out in detail. Thus, Ref \cite{z3} 
studied type I-heterotic duality matching for the $Z$-orbifold model of 
Ref \cite{Sagnotti}. There it was realized that although naively there is 
a discrepancy between the type I and heterotic spectra at the orbifold 
points, it disappears once orbifold singularities are blown-up and 
(anomalous) gauge symmetries are broken. The origin of this discrepancy is 
the following. If the type I orbifold model contains only $D9$-branes, then 
the open string sector gives rise to charged matter fields identical to 
those in the untwisted sector of the corresponding heterotic dual. The 
heterotic untwisted sector also contains gauge singlets which are geometric 
moduli. These are found in the untwisted closed string sector of the type I 
model. Since heterotic strings are 
closed strings, the heterotic orbifold model also contains twisted sectors. 
These give rise to certain singlets charged under $U(1)$ factors in the gauge 
group, but neutral under the non-Abelian symmetries. Their duals are found 
in the twisted closed string sectors of the type I model. The heterotic 
twisted sectors also contain matter fields charged under the non-Abelian 
gauge groups. These have no type I counterparts. This is precisely the 
descrepancy between the type I and heterotic spectra mentioned above.

{}The descrepancy is resolved by the fact that in the heterotic string model, 
there always is a {\em perturbative} superpotential such that after 
appropriate Higgsing the extra twisted matter fields (charged under 
non-Abelian gauge groups) become heavy and decouple from the massless 
spectrum. This Higgsing always involves blowing-up orbifold singularities, 
and can sometimes also involve breaking of (anomalous) gauge symmetries. 
Thus, in the $Z$-orbifold case \cite{Sagnotti} studied in Ref  \cite{z3} the 
relevent terms in the superpotential are {\em renormalizable} Yukawa 
couplings of the form $ST^2$ (this is only a symbolic way of writing these 
couplings as certain indices are suppressed), where $S$ are the orbifold 
blow-up modes, whereas $T$ are the extra twisted matter fields. In 
Ref \cite{z7} we worked out another chiral $N=1$ type I model on 
${\bf Z}_7$ orbifold. There the couplings responsible for making the extra 
twisted matter fields heavy were {\em non-renormalizable} couplings of the 
form $ST^2 Q^2$, where $Q$ stands for untwisted matter fields acquiring 
vevs in order to break anomalous $U(1)$ gauge symmetry present in that model.

{}In this paper we give a prescription (following from tadpole cancellation 
conditions given in Appendix \ref{tadpoles}) for contructing type I models 
with only $D9$-branes on general Abelian orbifolds (the latter have odd 
order or else $D5$-branes would be present). Constructing their heterotic 
duals is not difficult. For a given type I model on $T^{2d}/G$ symmetric 
Abelain orbifold (of odd order), one works out the Chan-Paton matrices 
$\gamma_g$ ($g\in G$) according to the prescription given in 
Appendix \ref{tadpoles}. Then the heterotic dual is constructed by 
compactifying the ${\mbox{Spin}}(32)/{\bf Z}_2$ heterotic string on 
$T^{2d}/G$ with gauge connection (shift) in each twisted sector (labeled by) 
$g$ given by $\gamma_g$. The duality matching always goes as described above. 
To illustrate the rules and duality matching, we consider type I 
compactification on ${\bf Z}_3\otimes {\bf Z}_3$ orbifold (which is not a 
${\bf Z}_N$ orbifold).

{}We also generalize the rules of Appendix \ref{tadpoles} to a certain class 
of orbifolds with both $D9$- and $D5$-branes (the orbifold has even order 
in this case). To illusterate the rules we construct a type I 
compactification on ${\bf Z}_6$ orbifold. The model has $N=1$ supersymmetry, 
$U(6)\otimes U(6)\otimes U(4)$ gauge group coming from the 99 strings, 
$U(6)\otimes U(6)\otimes U(4)$ gauge group coming from the 55 strings, and 
chiral matter coming from all three open string sectors (99, 55 and 59). 
This model has anomalous $U(1)$ gauge symmetry.

{}The paper is organized as follows. In section II we construct type I 
compactification on ${\bf Z}_3\otimes {\bf Z}_3$ orbifold. In section III 
we construct the heterotic dual. In section IV we give perturbative 
superpotentials for these models. In section V we discuss the moduli space, 
and explain the matching between type I and heterotic tree-level massless 
spectra. In section VI we construct type I compactification on ${\bf Z}_6$ 
orbifold. 
In section VII we give conclusions and remarks. 
Appendix \ref{tadpoles} contains the rules for constructing type I 
compactifications with 
only $D9$-branes on general Abelian orbifolds (of odd order). Appendix 
\ref{D5tadpoles} generalizes these rules to certain type I orbifolds 
(of even order) with both $D9$- and $D5$-branes.

\section{Type I String on ${\bf Z}_3 \otimes {\bf Z}_3$ Orbifold}

{}In this section we discuss the construction of the type I model on 
${\bf Z}_3 \otimes {\bf Z}_3$ orbifold.
Let us start from the type IIB string model compactified on the six-torus 
$T^2\otimes T^2\otimes T^2$, where each of the two-tori $T^2$ has 
a ${\bf Z}_3$ rotational symmetry. 
This model has $N=8$ supersymmetry. Let us now consider the 
symmetric 
${\bf Z}_3 \otimes {\bf Z}_3$ orbifold model generated by the twists
\begin{eqnarray}
 T_3 &=& (\theta,\theta,0 \vert\vert \theta, \theta, 0)~, \\
 T_3^{\prime} &=& (0, \theta,\theta \vert\vert 0, \theta, \theta )~.
\end{eqnarray} 
Here $\theta$ is a $2\pi /3$ rotation of a complex boson (we have complexified 
the six real bosons into three complex bosons). The double vertical 
line separates the right- and left-movers of the string. The resulting model 
has $N=2$ space-time supersymmetry. This model has the following moduli. 
There are 8 NS-NS fields $\phi,B_{\mu\nu},B_{i{\bar i}},g_{i{\bar i}}$, 
and 8 R-R fields 
$\phi^\prime,B^\prime_{\mu\nu},B^\prime_{i{\bar i}},
C^\prime_{\mu\nu i{\bar i}}$.

{}Let us now consider the orientifold projection of this model. The closed 
string sector (which is simply the subspace of the
Hilbert space of the original type IIB spectrum invariant under the orientifold
projection $\Omega$) contains the $N=1$ supergravity multiplet, and 
3 untwisted 
(the NS-NS fields that survive the $\Omega$ projection are 
$g_{i{\bar i}}$, whereas the R-R fields that are kept are 
$B^\prime_{i{\bar i}}$; note that the  
NS-NS field $\phi$ and the R-R field $B^\prime_{\mu\nu}$ also survive and, 
enter in the dilaton supermultiplet)
and 81 twisted chiral supermultiplets (which are neutral under the gauge 
group of the model). 
For consistency ({\em i.e.}, tadpole cancellation; see Appendix \ref{tadpoles}
for details), we must include 
the open string sector. Note that in this model we only have $D9$-branes 
but no $D5$-branes since the orbifold group does not contain an order 
two element. Thus, we only have 
$99$ open strings. The gauge group consistent with tadpole cancellation 
then is $U(4)\otimes U(4) \otimes U(4) \otimes SO(8)$. The $99$ open 
strings also 
give rise to the following chiral matter fields:
$(\overline{\bf 4}, {\bf 1}, {\bf 1}, {\bf 8}_v)(+1,0,0)_L$, 
$({\bf 1},\overline{\bf 4}, \overline{\bf 4}, {\bf 1})(0,-1,-1)_L$ and
$({\bf 6}, {\bf 1}, {\bf 1}, {\bf 1})(+2,0,0)_L$. In addition, there 
are fields that can be obtained by permuting the three $U(4)$'s (this 
permutation must be accompanied by changing the irrep of the second and
the third $U(4)$ to its complex conjugate).
Here the first four entries in bold font indicate the irreps of the
$SU(4)\otimes SU(4) \otimes SU(4) \otimes SO(8)$ subgroup, whereas the 
$U(1)^3$ charges  are given in the parenthesis. 
The subscript $L$ indicates the space-time helicity of the corresponding 
fermionic fields. The massless spectrum of this model is summarized in Table I.

{}Note that the $U(1)^3$ gauge symmetry is anomalous. We can form a linear
combination of these $U(1)$'s such that only one of them
is anomalous (this combination is given by $Q_1-Q_2-Q_3$, where $Q_{1,2,3}$ 
are the first, second and third $U(1)$ charges, respectively). The total 
$U(1)$ anomaly is $+36$. 
By the generalized Green-Schwarz
mechanism \cite{GS} some of the fields charged under $U(1)$ will acquire 
vevs to cancel the Fayet-Illiopoulos $D$-term.

\section{Heterotic String on ${\bf Z}_3 \otimes {\bf Z}_3$ Orbifold}

{}In this section we give the construction of
the heterotic string model that is (candidate) dual to the type I
model considered in the previous section. Let us start from the Narain model
with $N=4$ space-time supersymmetry in four dimensions. Let the momenta of the
internal (6 right-moving and 22 left-moving) world-sheet bosons span the 
(even self-dual) Narain lattice 
$\Gamma^{6,22}=(\Gamma^{2,2})^3\otimes\Gamma^{16}$.
Here $\Gamma^{16}$ is the ${\mbox{Spin}}(32)/{\bf Z}_2$ lattice, whereas the 
lattice $\Gamma^{2,2}$ (that corresponds to a two-torus $T^2$) is spanned by 
the momenta $(p_R \vert\vert p_L)$ with
\begin{eqnarray}
 p_{L,R}={1\over 2}m_i {\tilde e}^i \pm n^i e_i ~.
\end{eqnarray}
Here $m_i$ and $n^i$ are integers, $e_i \cdot e_j =g_{ij}$ is the constant 
background metric, and $e_i \cdot
{\tilde e}^j={\delta_i}^j$. Note that we could have included the constant
anti-symmetric background tensor field $B_{ij}$, but for now we will set it 
equal to zero.

{}This Narain model has the gauge group $SO(32) \otimes U(1)^6$. The first 
factor $SO(32)$ comes from the $\Gamma^{16}$ lattice (the $480$ roots of 
length squared 2), and 16 oscillator excitations of the corresponding 
world-sheet bosons (the latter being in the Cartan subalgebra of $SO(32)$).
The factor $U(1)^6$ comes from the oscillator excitations of the six 
left-moving world-sheet bosons corresponding to 
$\Gamma^{6,6}=(\Gamma^{2,2})^3$. Note that
there are also six additional vector bosons coming from the oscillator 
excitations of the right-moving world-sheet bosons corresponding to 
$\Gamma^{6,6}$. These vector bosons are part of the $N=4$ supergravity 
multiplet.

{}Next consider the ${\bf Z}_3 \otimes {\bf Z}_3$ orbifold model 
(with non-standard embedding
of the gauge connection) obtained via twisting the above Narain model by the
following ${\bf Z}_3$ twists:
\begin{eqnarray}
 T_3 &=& (\theta,\theta, 0 \vert\vert \theta,\theta,0 \vert
 ({1\over 3})^{4} (-{1\over 3})^4 0^8)~, \\
 T_3^{\prime} &=& (0,\theta,\theta \vert\vert 0,\theta,\theta \vert
 ({1\over 3})^{2} 0^2 (-{1\over 3})^2 0^2 ({1\over 3})^{2} (-{1\over 3})^2
  0^4)~. \\
\end{eqnarray}   
Here $\theta$ is a $2\pi /3$ rotation of a complex boson (we have complexified 
the original six real bosons into three complex ones). Thus, the first three
entries correspond to the ${\bf Z}_3$ twists of the three right-moving
complex bosons (coming from the six-torus).
The double vertical line separates the right- and left-movers.
The first three left-moving entries correspond to the ${\bf Z}_3$ twists of 
the three left-moving complex bosons (coming from the six-torus). The single 
vertical line separates the latter from the sixteen real bosons corresponding 
to the $\Gamma^{16}$ lattice. The latter are written in the $SO(32)$ basis. 
Thus, for example, $(+1, -1, 0^{14})$ is a root of $SO(32)$ with length 
squared two. There are $480$ roots like this in the $\Gamma^{16}$ lattice, 
and they are descendents of the identity irrep of $SO(32)$. The lattice 
$\Gamma^{16}$ also contains one of the spinor irreps as well. Thus, we will 
choose this spinor irrep to contain the momentum states of the form 
$(\pm {1\over2},...,\pm {1\over 2})$ with even number of plus signs.

{}Now we are ready to discuss the orbifold model generated by the twists
$T_3$ and $T_3^{\prime}$ . 
This model has $N=1$ space-time supersymmetry, and gauge group
$U(4)\otimes U(4) \otimes U(4) \otimes SO(8)$, the same as the type I model 
discussed in the previous 
section. The untwisted sector gives rise to the $N=1$ supergravity multiplet
coupled to the $N=1$ Yang-Mills gauge multiplet in the adjoint of 
$U(4)\otimes U(4) \otimes U(4) \otimes SO(8)$.
The matter fields in the untwisted sector are the 
same as those in the open string sector of the type I model.
There are also chiral multiplets
neutral under the gauge group: $3({\bf 1},{\bf 1},{\bf 1},{\bf 1})(0,0,0)_L$. 
Note that these contain
6 scalar fields that are the left-over geometric moduli whose vevs parametrize 
the moduli space $[SU(1,1,{\bf Z})\backslash SU(1,1)/ U(1)]^3$. 
(This is the subspace
of the original Narain moduli space 
$SO(6,6,{\bf Z})\backslash SO(6,6)/SO(6)\otimes 
SO(6)$ that is invariant under the twist.) Actually, the (perturbative) 
moduli space of this model is larger, and we will return to this point 
later on.

{}Next, consider the twisted sector. Since both $T_3$ and $T_3^{\prime}$
are of order $3$, their respective contributions to the one-loop partition
function can have a non-trivial relative ${\bf Z}_3$ phase between them
which we denote as $\phi(T_3,T_3^{\prime})$ 
({\em i.e.}, $\phi(T_3,T_3^{\prime})=0,1/3,2/3$). 
The states that survive the $T_3$ 
projection in the $T_3^{\prime}$ sector have $T_3$ phase
$\phi(T_3,T_3^{\prime})$, and the states that survive the $T_3$ projection
in the inverse twisted sector $(T_3^{\prime})^{-1}$ have $T_3$ phase
$-\phi(T_3,T_3^{\prime})$. String consistency requires that the states
that survive the $T_3^{\prime}$ projection in the $T_3$ sector must have
$T_3^{\prime}$ phase $-\phi(T_3,T_3^{\prime})$. Similarly, the states
that survive the $T_3^{\prime}$ projection in the inverse twisted sector
$(T_3)^{-1}$ must have $T_3^{\prime}$ phase $\phi(T_3,T_3^{\prime})$.
The twisted sector field content depends
on the relative phase $\phi(T_3,T_3^{\prime})$ and so it gives rise to three
different heterotic string models. It turns out that the model with 
$\phi(T_3,T_3^{\prime})=2/3$ is dual to the type I model described in the
previous section. Here, we list out the twisted sector field content of this
model. We have the following chiral
supermultiplets: $9({\bf 1},{\bf 1},{\bf 1},{\bf 1})(\pm 4/3, \pm 4/3,0)_L$,
$9({\bf 1},{\bf 6},{\bf 1},{\bf 1})(+4/3,-2/3,0)$ and 
$9({\bf 6},{\bf 1},{\bf 1},{\bf 1})(+2/3,-4/3,0)$ together with 
fields obtained by permuting the three $U(4)$'s (this permutation must 
be accompanied by changing the irrep of the second and the third $U(4)$ 
to its complex conjugate). In addition, there are 
$27({\bf 1},{\bf 1},{\bf 1},{\bf 1})(-4/3,+4/3,+4/3)_L$ and
$27({\bf 1},{\bf 1},{\bf 1},{\bf 8_s})(+2/3,-2/3,-2/3)_L$
coming from $T_3 + T_3^{\prime}$ sector (and its conjugate).
Here we note that the factors ``9'' and ``27'' come from the number of 
fixed points in different sectors of the 
${\bf Z}_3 \times {\bf Z}_3$ orbifold we are considering. The $27$
singlets and the $27$ ${\bf 8_s}$ of $SO(8)$ are all projected out in
the other two models ($\phi(T_3,T_3^{\prime})=0,1/3$), and hence
their spectra do not match with the type I model. 

{}We summarize the massless spectrum of the heterotic string model
which is the candidate dual of the type I model in Table II.
Note that the $U(1)^3$ gauge symmetry is anomalous. Again, only one linear
combination of the three $U(1)$'s is anomalous. Thus, the contributions of the 
untwisted and twisted sectors into the trace anomaly are
$+36$ and $27 \times (+36)$, respectively,
so that the total trace anomaly is $+972$. By the generalized Green-Schwarz 
mechanism \cite{GS} some of the fields charged under $U(1)$ will acquire 
vevs to cancel the Fayet-Illiopoulos $D$-term. 

\section{Superpotential}

{}In this section we discuss the perturbative superpotentials for the type I 
and heterotic string models discussed in the previous sections. Studying the 
couplings and flat directions in these superpotentials will enable us to 
make the type I-heterotic duality map more precise.

{}Let us start from the type I model of section II.
We refer the reader to Table I for the massless spectum as well as
our notation. 
Note that perturbatively the 
81 chiral singlets coming from the closed string sector are flat. 
This can be explicitly seen by computing the scattering amplitudes for 
these modes within the framework of the conformal field theory of 
orbifolds \cite{DFMS}. On the 
other hand, the matter fields coming from the $99$ open string sector 
have three (and, of course, some higher) point couplings. The lowest order 
superpotential can be written as (the calculation of the type I 
superpotential is completely analogous to that of the heterotic one in the 
untwisted sector)
\begin{equation}
 W_I =  \lambda\epsilon_{abc} {\mbox{Tr}} (P_a P_b Q_c) +...~.
\end{equation}
Due to the presence of the anomalous $U(1)$, some of the fields that are 
charged under this $U(1)$ 
must acquire vevs to cancel the Fayet-Illiopoulos $D$-term. 
This results in breakdown of gauge symmetry, yet the space-time 
supersymmetry is preserved.

{}Now let us turn to the heterotic string model. The superpotential of this 
model is more involved than that of the type I model as there are non-trivial
couplings between the twisted sector fields. 
The superpotential for the 
heterotic string model thus reads (here we are only interested in the 
general structure of the {\em non-vanishing} terms):
\begin{eqnarray}
 W_H =&&\lambda^{\prime} \epsilon_{abc} {\mbox{Tr}} (P_a P_b Q_c)
       +\Lambda^{(\alpha \alpha^{\prime} \alpha^{\prime \prime})
                 (\beta \beta^{\prime} \beta^{\prime \prime})
                 (\gamma \gamma^{\prime} \gamma^{\prime \prime})}
         {\mbox{Tr}} (S_{\alpha \beta \gamma} 
                      T_{\alpha^{\prime} \beta^{\prime} \gamma^{\prime}}
       T_{\alpha^{\prime \prime} \beta^{\prime \prime}\gamma^{\prime\prime}} )
\nonumber\\
      &+& \Lambda_1^{(\alpha \alpha^{\prime} \alpha^{\prime \prime})
                  (\beta \beta^{\prime})(\gamma \gamma^{\prime \prime})}
           {\mbox{Tr}} (S_{\alpha \beta \gamma} 
                        T^{1+}_{\alpha^{\prime} \beta^{\prime}}
                        T^{3+}_{\alpha^{\prime \prime} \gamma^{\prime\prime}} )
      + \Lambda_2^{(\alpha \alpha^{\prime})
                   (\beta \beta^{\prime} \beta^{\prime \prime})
                   (\gamma \gamma^{\prime \prime})}
           {\mbox{Tr}} ( S_{\alpha \beta \gamma} 
                         T^{1-}_{\alpha^{\prime} \beta^{\prime}}
                         T^{2-}_{\beta^{\prime \prime} \gamma^{\prime\prime}} )
\nonumber \\
&+& \Lambda_3^{(\alpha \alpha^{\prime \prime})
                   (\beta \beta^{\prime})
                   (\gamma \gamma^{\prime} \gamma^{\prime \prime})}
           {\mbox{Tr}} ( S_{\alpha \beta \gamma} 
                         T^{2+}_{\beta^{\prime} \gamma^{\prime}}
                         T^{3-}_{\alpha^{\prime \prime}\gamma^{\prime\prime}} )
+...~.
\end{eqnarray}
(The notation for the fields is given in Table II.) The couplings
$\Lambda$, $\Lambda_1$, $\Lambda_2$ and $\Lambda_3$
are non-vanishing if the orbifold {\em {space group}} selection rules are
satisifed. These rules read as follows. 
$\Lambda^{(\alpha \alpha^{\prime} \alpha^{\prime \prime})
                 (\beta \beta^{\prime} \beta^{\prime \prime})
                 (\gamma \gamma^{\prime} \gamma^{\prime \prime})}\not=0$ if 
and only if $\alpha=\alpha^\prime=\alpha^{\prime\prime}$ or 
$\alpha\not =\alpha^\prime\not=\alpha^{\prime\prime}\not=\alpha$, and 
similarly for the $\beta$- and $\gamma$-indices. 
$\Lambda_1^{(\alpha \alpha^{\prime} \alpha^{\prime \prime})
                 (\beta \beta^{\prime})(\gamma \gamma^{\prime \prime})}\not=0$
if and only if $\alpha=\alpha^\prime=\alpha^{\prime\prime}$ or 
$\alpha\not =\alpha^\prime\not=\alpha^{\prime\prime}\not=\alpha$, 
$\beta=\beta^\prime$ and $\gamma=\gamma^{\prime\prime}$. The selection rules 
for the $\Lambda_2$ and $\Lambda_3$ couplings are similar to those for the 
$\Lambda_1$ couplings. 
Here we note that, say, couplings 
$\Lambda^{(\alpha \alpha^{\prime} \alpha^{\prime \prime})
                 (\beta \beta^{\prime} \beta^{\prime \prime})
                 (\gamma \gamma^{\prime} \gamma^{\prime \prime})}$ with 
$\alpha\not =\alpha^\prime\not=\alpha^{\prime\prime}\not=\alpha$, and 
similarly for the $\beta$- and $\gamma$-indices,
are exponentially suppressed in the limit of large volume 
compactification, whereas the couplings with
$\alpha=\alpha^{\prime}=\alpha^{\prime \prime}$,
$\beta=\beta^{\prime}=\beta^{\prime \prime}$ and 
$\gamma=\gamma^{\prime}=\gamma^{\prime \prime}$  are not suppressed.
This is because in the former case, the corresponding fields are coming from  
different fixed points so that upon taking them apart 
(in the limit of large volume of the orbifold) their coupling becomes 
weaker and weaker. Similar statements hold for the $\Lambda_1$, $\Lambda_2$ 
and $\Lambda_3$ couplings as well.

{}Notice that upon the singlets $S_{\alpha \beta \gamma}$ (which are  
27 blow-up modes) 
acquiring vevs, the states $T_{\alpha \beta \gamma}$ and 
$T^{a \pm}_{\alpha \beta}$
become heavy and decouple from the massless spectrum. 
Thus, after blowing up the orbifold singularities on the heterotic side 
(combined with some of the untwisted charged matter fields acquiring vevs 
to cancel the $D$-term), we can match the massless spectrum to that of the 
type I model (where the charged matter must acquire vevs to cancel the effect 
of the anomalous $U(1)$). Note the crucial role of the perturbative 
superpotential in this matching. It is precisely such that all the extra 
fields on the heterotic side can be made massive by blowing up the orbifold.

\section{Moduli Space}

{}We now turn to the discussion of the moduli spaces for the type I and 
heterotic models considered in the previous sections. Let us start with 
the heterotic model. The (perturbative) moduli space of the corresponding 
Narain model before orbifolding is 
$SO(6,22,{\bf Z}) \backslash SO(6,22) /SO(6)\otimes SO(22)$. After 
orbifolding we have two types of moduli: those coming from the untwisted 
sector, and those coming from the twisted sector. The untwisted sector 
moduli parametrize the coset 
$[SU(1,3,{\bf Z}) \backslash SU(1,3)/SU(3) \otimes U(1)]^3$. 
The subspace $[SU(1,1,{\bf Z}) \backslash SU(1,1)/ U(1)]^3$ of this 
moduli space is parametrized by $6$ neutral singlets $\phi_{a}$ that 
correspond to the 
left-over geometric moduli (coming from the constant metric $g_{ij}$ and 
antisymmetric tensor $B_{ij}$ fields). The other $12$ moduli correspond to 
the flat directions in the superpotential for the fields $P_a$, $Q_a$
and $\Phi_a$. 
(These are the left-over moduli coming from the $6\times 16$ Wilson lines 
$A^I_i$, $I=1,...,16$, in the original Narain model.)

{}Next, we turn to the twisted moduli of the heterotic string model. 
In the twisted sectors, we have the chiral superfields 
$S^{a \pm}_{\alpha \beta}$, $T^{a \pm}_{\alpha \beta}$,
$S_{\alpha \beta \gamma}$ and $T_{\alpha \beta \gamma}$. 
At a generic point (upon giving appropriate vevs to the 
$S_{\alpha \beta \gamma}$ fields), the fields $T^{a \pm}_{\alpha \beta}$
and $T_{\alpha \beta \gamma}$ become massive (according to the couplings 
in the superpotential) and there is no superpotential for
the singlets $S^{a \pm}_{\alpha \beta}$ and $S_{\alpha \beta \gamma}$
(the blow-up modes of the ${\bf Z}_3 \otimes {\bf Z}_3$ orbifold).
The singlets $S^{a \pm}_{\alpha \beta}$ are not charged under the
anomalous $U(1)$. Hence, all of them survive the Higgsing process
and they match with the corresponding singlets in the type I spectrum.
However, the fields $S_{\alpha\beta\gamma}$ are charged under the anomalous
$U(1)$. Therefore, a linear combination of them will be eaten in the 
super-Higgs mechanism. So, naively, not all the $27$ chiral fields 
survive after Higgsing. This would pose a problem for matching of the type I 
and heterotic models. Note, however, that the type I model has anomalous 
$U(1)$, and to cancel the $D$-term one needs to give vevs to the 
corresponding charged fields. So, generically, the fields $P_a$, $Q_a$ 
and $\Phi_a$ will acquire vevs (on the type I side) to break the anomalous 
$U(1)$. Thus, to match the type I and heterotic models we have to give vevs 
to the $P_a$, $Q_a$ and $\Phi_a$ fields on the heterotic side as well. 
Then, we will have $27$ neutral chiral superfields (which, after Higgsing, 
are a mixture of the original fields $S_{\alpha\beta\gamma}$ and the 
untwisted matter fields) on the heterotic side, which do correspond to the 
$27$ neutral chiral superfields coming from the twisted closed string sector 
of the type I model. Thus, the matching is complete after giving appropriate 
vevs to {\em both} twisted {\em and} untwisted fields on the heterotic side, 
as well as giving appropriate vevs to open string sector matter fields, and 
$27$ twisted closed string moduli. 
Upon
breaking the anomalous $U(1)$, the dilaton may mix with other gauge 
singlets \cite{DSW}.
{\em A priori}, the mixing is different on the type I and the heterotic
side. To 
make the matching precise, one generically has to appropriately tune the 
dilaton plus $\phi_{a}$ geometric moduli on both sides.

{}Thus, the moduli spaces (at generic points) of both type I and heterotic 
models are the same (at least at tree-level). They are described by the 
untwisted moduli of the heterotic string, or equivalently, the moduli 
coming from the untwisted closed string sector and the open string sector 
of the type I model (these parametrize the coset 
$[SU(1,3,{\bf Z}) \backslash SU(1,3)/SU(3) \otimes U(1)]^3$), plus the 
$2\times 81$ twisted moduli in the heterotic string model, or equivalently, 
the moduli coming from the twisted closed string sector of the type I model. 
The (perturbative) moduli space (of the heterotic model) is schematically 
depicted in Fig.1.

{}It is worth noticing the role of anomalous $U(1)$ in $N=1$ type I-heterotic
duality. To cancel the Fayet-Illiopoulos $D$-term, fields that are charged 
under the anomalous $U(1)$ will generically acquire vevs. As a result,
the extra twisted matter fields
in the heterotic model are higgsed away and the matching of the
massless spectra of the type I and heterotic models is precise.
The appearance of massless twisted matter fields $T_{\alpha \beta \gamma}$
and $T^{a \pm}_{\alpha \beta}$ on 
the heterotic side is a perturbative effect. On the type I side this 
effect is non-perturbative, and reflects the fact that from type I point 
of view there is a (non-perturbative) singularity 
in the moduli space (or, more precisely, 
a singular subspace of the full moduli space).
Notice that the fields $T_{\alpha \beta \gamma}$
and $T^{a \pm}_{\alpha \beta}$ in the heterotic model get heavy 
via couplings in the {\em perturbative} 
superpotential. This indicates the importance
of perturbative superpotential in $N=1$ type I-heterotic duality.

\section{Type I String on ${\bf Z}_6$ Orbifold}

{}In this section we discuss the compactification of type I strings on 
${\bf Z}_6$ orbifold. Since ${\bf Z}_6$ has an order two element, we
have to include both $D9$- and $D5$-branes. Let us start from the 
type IIB string model compactified on the six-torus 
$T^2\otimes T^2\otimes T^2$, where each of the two-tori has 
a ${\bf Z}_3$ and a ${\bf Z}_2$ rotational symmetry. 
This model has $N=8$ supersymmetry.
Let us now consider the 
symmetric 
${\bf Z}_6$ orbifold model generated by the twists
\begin{eqnarray}
 T_3 &=& (\theta,\theta,\theta \vert\vert \theta, \theta, \theta)~, \\
 T_2 &=& (\sigma, \sigma,0 \vert\vert \sigma, \sigma, 0 )~.
\end{eqnarray} 
Here $\theta$ is a $2\pi /3$ rotation of a complex boson (we have complexified 
the six real bosons into three complex bosons). Similarly, $\sigma$ is
a $\pi$ rotation of a complex boson. The double vertical 
line separates the right- and left-movers of the string. The resulting model 
has $N=2$ space-time supersymmetry. This model has the following moduli. 
This model has the following moduli. 
There are 12 NS-NS fields $\phi,B_{\mu\nu},B_{i{\bar j}},g_{i{\bar j}}$ 
($i,j=1,2$), $B_{3{\bar 3}},g_{3{\bar 3}}$
and 12 R-R fields 
$\phi^\prime,B^\prime_{\mu\nu},B^\prime_{i{\bar j}},
C^\prime_{\mu\nu i{\bar j}}$ ($i,j=1,2$), $B^\prime_{3{\bar 3}},
C^\prime_{\mu\nu 3{\bar 3}}$.

{}Let us now consider the orientifold projection of this model. The closed 
string sector (which is simply the subspace of the
Hilbert space of the original type IIB spectrum invariant under the orientifold
projection $\Omega$) contains the $N=1$ supergravity multiplet, and 
5 untwisted 
(the NS-NS fields that survive the $\Omega$ projection are 
$g_{i{\bar j}}$ ($i,j=1,2$) and $g_{3{\bar 3}}$, whereas the R-R fields that 
are kept are 
$B^\prime_{i{\bar j}}$ ($i,j=1,2$) and $B^\prime_{3{\bar 3}}$; note that the  
NS-NS field $\phi$ and the R-R field $B^\prime_{\mu\nu}$ also survive and, 
enter in the dilaton supermultiplet)
and 29 twisted chiral 
supermultiplets (which are neutral under the gauge 
group of the model). 
For consistency ({\em i.e.}, tadpole cancellation; see 
Appendix \ref{D5tadpoles} for details), we must include 
the open string sector. In this model, there are both $D9$- and $D5$-branes.
The gauge group consistent with tadpole cancellation 
then is $[U(6)\otimes U(6) \otimes U(4)]^2$. The gauge bosons
of the first $U(6) \otimes U(6) \otimes U(4)$ factor come from the $99$
sector whereas the gauge bosons of the second $U(6) \otimes U(6) \otimes U(4)$
come from the $55$ sector. The $99$ sector also gives rise to  
the following chiral matter fields charged under the first 
$U(6) \otimes U(6) \otimes U(4)$:
$2({\bf 15},{\bf 1},{\bf 1})(+2,0,0)_L$,
$2({\bf 1},\overline{\bf 15},{\bf 1})(0,-2,0)_L$,
$2(\overline{\bf 6},{\bf 1},\overline{\bf 4})(-1,0,-1)_L$,
$2({\bf 1},{\bf 6},{\bf 4})(0,+1,+1)_L$,
$({\bf 6},\overline{\bf 6},{\bf 1})(+1,-1,0)_L$,
$(\overline{\bf 6},{\bf 1},{\bf 4})(-1,0,+1)_L$ and
$({\bf 1},{\bf 6},\overline{\bf 4})(0,+1,-1)_L$.
Here the first three entries in bold font indicate the irreps of the
$SU(6)\otimes SU(6) \otimes SU(4)$ subgroup, whereas the 
$U(1)^3$ charges  are given in the parenthesis. 
The subscript $L$ indicates the space-time helicity of the corresponding 
fermionic fields. The $55$ sector gives rise to similar
chiral matter fields but charged under the second 
$U(6) \otimes U(6) \otimes U(4)$. 
The fact that the $99$ and $55$ sector have the same spectrum 
follows from $T$-duality.
The $59$ sector provides the chiral matter fields charged under both $99$ and 
$55$ gauge groups. The massless spectrum of this model is summarized in 
Table III.  Note that this model has anomalous $U(1)$.

{}The heterotic dual of this model would correspond to a non-perturbative
vacuum with small instantons. The $55$ sector gauge group cannot be
Higgsed away completely (due to the presence of a tree-level
superpotential), but an $SU(2)$ gauge group (with no charged matter)
remains. Thus, the type I model constructed in this section is an
example of a non-perturbative chiral $N=1$ vacuum in four dimensions
from the heterotic viewpoint.

\section{Conclusions}

{}In this paper we discussed the rules for type I compactifications with 
only $D9$-branes on general Abelian orbifolds (of odd order), and gave the 
prescription for constructing their heterotic duals. We discussed the issues 
involved in the duality matching, and illustrated them on a particular 
example of ${\bf Z}_3\otimes {\bf Z}_3$ orbifold. We also generalized the 
rules for type I model building to certain cases with both $D9$- and 
$D5$-branes present. The example of ${\bf Z}_6$ orbifold we consider in this 
paper would be an interesting arena for testing the validity of 
type I-heterotic duality. 

\acknowledgements

{}We would like to thank Ignatios Antoniadis, Alex Buchel, Tom Taylor and 
Cumrun Vafa
for discussions.
The research of G.S. was partially supported by the 
National Science Foundation. G.S. would also like to thank
Joyce M. Kuok Foundation for financial support.
The work of Z.K. was supported in part by the grant NSF PHY-96-02074, 
and the DOE 1994 OJI award. 
Z.K. would also like to thank Albert and Ribena Yu for 
financial support.

\appendix
\section{Tadpoles for $D9$-branes}\label{tadpoles}

{}In this appendix we discuss the tadpole cancellation constraints for 
orbifold compactifications of type I strings without $D5$-branes.
In particular, we confine our attention to general Abelian 
${\bf Z}_{n_1} \otimes {\bf Z}_{n_2} \otimes \cdots \otimes {\bf Z}_{n_k}$
orbifolds
where $n_1,n_2,\dots,n_k$ are all odd integers (if the orbifold group
contains an order two element $R$, then the sector $\Omega R$
would contain $D5$-branes).
The constraints for orbifold compactifications with $D5$-branes will be
given in the next section.

{}There are two kinds of constraints we need to consider. The first one
comes from the cancellation of the untwisted tadpoles for the $D9$-branes. 
This constraint is the same in all dimensions and leads to the statement 
that there are $16$ $D9$-branes {\em not} counting the orientifold images. 
(This last statement is only correct if the NS-NS antisymmetric background 
$B_{ij}$ is set equal to zero \cite{Sagnotti1}; see below.) The other 
constraint 
comes from the cancellation of the twisted tadpoles for the $D9$-branes. 
The twisted tadpoles have been computed for certain cases in six dimensions 
\cite{GP,GJ}, and in four dimensions \cite{Sagnotti,BL}.
The cases of ${\bf Z}_{N}$ orbifolds (odd prime $N$) in general $D$
dimensions have been discussed in \cite{z7}. Here, we generalize these
tadpole cancellation conditions to general Abelian orbifolds with no order 
two element.

{}Consider compactification on $T^{2d}/G$ where 
$G={\bf Z}_{n_1} \otimes {\bf Z}_{n_2} \otimes \cdots \otimes {\bf Z}_{n_k}$
for odd integers $n_1, n_2, \dots, n_k$.
Let $g_1, g_2, \dots , g_k$ be the generators of ${\bf Z}_{n_1}, 
{\bf Z}_{n_2}, \dots ,{\bf Z}_{n_k}$ respectively. The corresponding
twists are given by
\begin{equation}
 T_{g_a}=(t_1^a,t_2^a,...,t_d^a \vert\vert t_1^a,t_2^a,...,t_d^a)~.
\end{equation}
Here  $t_i^a$ are fractional numbers taking values in 
$\{0,1/n_a,2/n_a,...,(n_a-1)/n_a\}$. A given $t_i^a$ corresponds to a twist 
of the 
$i$-th complex boson by a $2\pi t_i^a$ rotation. (We have complexified the 
$2d$ real bosons into $d$ complex bosons.) The double vertical line separates 
the right- and left-movers of the string. 
Because we are considering symmetric orbifold, the right- and 
left-moving twists are the same. The elements 
$g(\alpha)$ of the the orbifold group $G$ can be written as 
$g(\alpha)=g_1^{\alpha_1} g_2^{\alpha_2} \cdots g_k^{\alpha_k}$, where 
$\alpha_a=0,1,...,n_a -1$. The corresponding twist $T_{g(\alpha)}$
is defined by $t_i (\alpha)=\sum_{a} \alpha_a t^a_i~({\mbox{mod}}~1)$, so 
that $0\leq t_i (\alpha)<1$. We note that the consistency of the orbifold 
requires that for each element $g(\alpha)$ the expression  
\begin{equation}\label{fixed}
   \prod_{i=1}^{d} 4\sin^2(\pi t_i (\alpha))~,
\end{equation}
where the factors with $t_i (\alpha)=0$ are not included in the product, be an
integer. In fact the latter is nothing but the number of fixed 
points (tori) in the $T_{g(\alpha)}$ twisted sector.

{}The orbifold action on Chan-Paton factors is described by the unitary 
matrices $\gamma_{g(\alpha)}$ that act on the string end-points.   
In our case $\gamma_{g(\alpha)}$ 
is a $16\times 16$ matrix (note that it is {\em not} a 
$32\times 32$ matrix because we have chosen {\em not} to count the 
orientifold images of the $D9$-branes). Since the orbifold
is Abelian, we can simultaneously diagonalize the matrices 
$\gamma_{g(\alpha)}$ for each ${g(\alpha)} \in G$. Then, the most general 
form of $\gamma_{g(\alpha)}$
is given by
\begin{equation}
\gamma_{g(\alpha)} = {\mbox{diag}} 
 ([\omega (\alpha)]^{\ell_1 (\alpha)},[\omega (\alpha)]^{\ell_2 (\alpha)]},
\dots,
 [\omega (\alpha)]^{\ell_{16} (\alpha)} ) ~,
\end{equation} 
where $\omega (\alpha)\equiv\exp(2\pi i /N (\alpha))$. Here $N (\alpha)$ is 
the order of ${g(\alpha)}$, {\em i.e.}, the smallest integer such that 
$N(\alpha) t_i (\alpha) \in {\bf Z}$ for 
$i=1,\dots,d$.

{}The twisted tadpole cancellation conditions are thus
\begin{equation}
 {\mbox{Tr}}(\gamma_{g(\alpha)})=16 p(\alpha)~,~~~p(\alpha)\equiv 
      \prod_{i=1}^{d} (-1)^{N (\alpha) t_i (\alpha)} \cos(\pi t_i (\alpha))
\end{equation}
for each ${g(\alpha)} \in G$.

{}Let us illustrate the above constraint by 
the ${\bf Z}_3 \otimes {\bf Z}_3$ orbifold considered in this paper:
\begin{eqnarray}
 T_{g_1} &=& (1/3,1/3,0 \vert\vert 1/3, 1/3, 0)~, \\
 T_{g_2} &=& (0, 1/3,1/3 \vert\vert 0, 1/3, 1/3 )~.
\end{eqnarray} 
The tadpole cancellation conditions read:
\begin{eqnarray}
 {\mbox{Tr}}(\gamma_{g_1}) &=& 4 \nonumber \\
 {\mbox{Tr}}(\gamma_{g_2}) &=& 4 \nonumber \\
 {\mbox{Tr}}(\gamma_{g_1 g_2^2}) &=& 4 \nonumber \\
 {\mbox{Tr}}(\gamma_{g_1 g_2}) &=& -2 ~.
\end{eqnarray}
The conditions for all other $g \in G$ can be derived from the ones above.
The solutions for $\gamma_{g_1}$ and $\gamma_{g_2}$ are then:
\begin{eqnarray}
\gamma_{g_1} &=&  {\bf I}_8 \otimes
                 \omega {\bf I}_{4} \otimes
                 \omega^2 {\bf I}_{4} \nonumber \\
\gamma_{g_2} &=&  {\bf I}_4 \otimes
                 \omega {\bf I}_{2} \otimes
                 \omega^2 {\bf I}_{2} \otimes
                  {\bf I}_2 \otimes
                 \omega {\bf I}_{2} \otimes
                 {\bf I}_2 \otimes
                 \omega^2 {\bf I}_{2}
\end{eqnarray}
where ${\bf I}_n$ denotes the $n \times n$ unit matrix. Thus, the gauge 
group is $U(4)^3 \otimes SO(8)$.

{}Here we also list a few non-supersymmetric models (that have never been 
discussed previously to the best of our knowledge):\\
$\bullet$ 6D ${\bf Z}_3\otimes {\bf Z}_3$ orbifold. The twists read:
\begin{eqnarray}
 T_{g_1}&=&(1/3,0\vert\vert 1/3,0)~,\\
 T_{g_2}&=&(0,1/3\vert\vert 0,1/3)~.
\end{eqnarray}
The gauge group is $U(8)\otimes U(8)$.\\
$\bullet$ 4D ${\bf Z}_3\otimes {\bf Z}_3$ orbifold. The twists read:
\begin{eqnarray}
 T_{g_1}&=&(1/3,0,0\vert\vert 1/3,0,0)~,\\
 T_{g_2}&=&(0,1/3,1/3\vert\vert 0,1/3,1/3)~.
\end{eqnarray}
The gauge group is $U(4)\otimes U(4)\otimes U(8)$.\\
$\bullet$ 4D ${\bf Z}_3\otimes {\bf Z}_3 \otimes {\bf Z}_3$ orbifold. The 
twists read:
\begin{eqnarray}
 T_{g_1}&=&(1/3,0,0\vert\vert 1/3,0,0)~,\\
 T_{g_2}&=&(0,1/3,0\vert\vert 0,1/3,0)~,\\
 T_{g_2}&=&(0,0,1/3\vert\vert 0,0,1/3)~.
\end{eqnarray}
The gauge group is $U(4)^4$.\\
$\bullet$ 4D ${\bf Z}_3\otimes {\bf Z}_5$ orbifold. The twists read:
\begin{eqnarray}
 T_{g_1}&=&(1/3,0,0\vert\vert 1/3,0,0)~,\\
 T_{g_2}&=&(0,1/5,2/5\vert\vert 0,1/5,2/5)~.
\end{eqnarray}
The gauge group is $U(4)^4$.\\
$\bullet$ 4D ${\bf Z}_9$ orbifold. The twist reads:
\begin{eqnarray}
 T_{g}&=&(1/9,2/9,4/9\vert\vert 1/9,2/9,4/9)~.
\end{eqnarray}
The gauge group is $U(4)^4$.

{}Finally, we would like to consider the cases with non-zero NS-NS 
antisymmetric background
$B_{ij}$. Although there are no massless scalars corresponding to these 
in type I theory (recall that there $B_{ij}$ fields are projected out of 
the spectrum after orientifolding), $B_{ij}$ can have certain quantized 
values. 
The quantization is due to the fact that to have a consistent orientifold the 
corresponding type IIB spectrum must be left-right symmetric. At generic 
values of $B_{ij}$ this symmetry is destroyed. There are, however, certain 
discrete $B_{ij}$ backgrounds compatible with the orientifold 
projection \cite{Sagnotti1}. The effect of non-zero $B_{ij}$ background 
is that the rank of the gauge group coming from the $SO(32)$ ({\em i.e.}, 
Chan-Paton) factor is reduced, depending on the rank $r$ (which is always 
even) of the matrix $B_{ij}$. That is, the number of the $D9$-branes 
required by the tadpole cancellation condition is no longer $16$ but 
$16/2^{r/2}$. All of 
the above formulas then get modified in the presence of rank $r$ $B_{ij}$ 
in an obvious way via replacing the factor $16$ everywhere by $16/2^{r/2}$.

\section{Tadpoles for $D9$- and $D5$-branes}\label{D5tadpoles}

{}In this section we generalize the tadpole cancellation conditions discussed 
in Appendix \ref{tadpoles} to the cases with both $D9$- and $D5$-branes 
present. In order to have $D5$-branes, the orbifold group $G$ must contain 
an order two element $R$. In general, there may be more than one order two 
elements $R_i$ in the orbifold group $G$. Then there will be corresponding 
$D5_i$-branes for each of these elements. Once the case with only one order 
two element is understood, the generalization to more general cases is 
relatively straightforward.

{}Let us, therefore, concentrate on the case where we have only one order 
two element $R\in G$. In fact, for now let us take the orbifold group 
$G={\bf Z}_{2N} \sim {\bf Z}_2 \otimes {\bf Z}_N$,where $N$ is odd. Note 
that $R\in {\bf Z}_2$. We can write the group elements as 
$G=\{ 1, g^1,...,g^{N-1},R,Rg^1,...,Rg^{N-1} \}=\{(Rg)^k,k=0,...,2N-1\}$ 
(note that $R^2=1$). The orientifold group element $\Omega$ gives rise to 
$D9$-branes, whereas the element $\Omega R$ gives rise to $D5$-branes. Let 
us denote the open string Chan-Paton matrices as $\gamma_k$  for the 
$D9$-branes, and as ${\tilde \gamma}_k$ for the $D5$-branes. Here 
$k=0,...,2N-1$, and the matrices with index $k$ even correspond to the $g^k$
twists, whereas those with the index $k$ odd correspond to the $Rg^k$ twists. 
Note that the untwisted Chan-Paton matrices are given by
$\gamma_0={\tilde \gamma}_0={\bf I}_{16}$, where ${\bf I}_{16}$ is a 
$16\times 16$ identity matrix. This latter fact follows from the untwisted 
tadpole cancellation conditions that require presence of 16 $D9$-branes and 
16 $D5$-branes (we are not counting the orientifold images here). 

{}Next, we need to understand the twisted tadpole cancellation conditions. 
Here we will write the general form of these conditions. The contributions 
to the twisted tadpoles come from the Klein bottle ${\cal K}$, cylinder 
${\cal C}$ and M{\"o}bius strip ${\cal M}$ (the total twisted tadpole is 
the sum ${\cal T}={\cal K}+{\cal C}+{\cal M}$): 
\begin{eqnarray}
 {\cal K}&=& {\sum_{k=1}^{2N-1}} {\cal K}_k ~,\\
 {\cal C}&=&\sum_{k=1}^{2N-1} (a_k  
 { \mbox{Tr}}(\gamma_{k})^2-2{\mbox{Tr}}(\gamma_{k}){\mbox{Tr}}
({\tilde \gamma}_{k})+
 {\tilde a}_k  { \mbox{Tr}}({\tilde \gamma}_{k})^2)~,\\
 {\cal M}&=&{\sum_{k=1}^{2N-1}} (b_k  
 { \mbox{Tr}}(\gamma^{-1}_{\Omega_k} \gamma^{T}_{\Omega_k})+
 {\tilde b}_k  { \mbox{Tr}}({\tilde \gamma}^{-1}_{\Omega_k} 
{\tilde \gamma}^{T}_{\Omega_k}))~.
\end{eqnarray}
Here $\Omega_k \equiv \Omega (Rg)^k$. Note that in the Klein bottle 
${\cal K}$ and M{\"o}bius strip ${\cal M}$ tadpoles we must {\em not} 
include the terms with $k=N$.

{}The Klein bottle contributions ${\cal K}_k$, as well as the coefficients 
$a_k,{\tilde a}_k,b_k,{\tilde b}_k$, are given by certain numerical factors 
independent of the Chan-Paton matrices. Note that according to the 
composition algebra for the matrices $\gamma_{\Omega_k}$ and 
${\tilde \gamma}_{\Omega_k}$ we have: 
\begin{eqnarray}
 { \mbox{Tr}}(\gamma^{-1}_{\Omega_k} \gamma^{T}_{\Omega_k})&=&\epsilon_k 
 { \mbox{Tr}}(\gamma_{2k})~,\\
 { \mbox{Tr}}({\tilde \gamma}^{-1}_{\Omega_k} {\tilde \gamma}^{T}_{\Omega_k})
&=&{\tilde 
   \epsilon}_k 
 { \mbox{Tr}}({\tilde \gamma}_{2k})~,
\end{eqnarray}
where $\epsilon_k$ and ${\tilde \epsilon}_k$ take values $\pm1$. Next note 
that since $(Rg)^{k+2N}=(Rg)^{k}$, we can rewrite the total twisted tadpole 
${\cal T}$ as a sum of two
pieces 
${\cal T}={\cal T}_{\mbox{{\small odd}}}+{\cal T}_{\mbox{{\small even}}}$, 
where
\begin{eqnarray}
 {\cal T}_{\mbox{{\small odd}}} &=&\sum_{k=1}^{N} (a_{2k-1}  
 { \mbox{Tr}}(\gamma_{2k-1})^2-2{\mbox{Tr}}(\gamma_{2k-1}){\mbox{Tr}}
({\tilde   
 \gamma}_{2k-1})+
 {\tilde a}_{2k-1}  { \mbox{Tr}}({\tilde \gamma}_{2k-1})^2)~,\\
 {\cal T}_{\mbox{{\small even}}}&=&\sum_{k=1}^{N-1} (a_{2k}  
 { \mbox{Tr}}(\gamma_{2k})^2-2{\mbox{Tr}}(\gamma_{2k}){\mbox{Tr}}({\tilde   
 \gamma}_{2k})+
 {\tilde a}_{2k}  { \mbox{Tr}}({\tilde \gamma}_{2k})^2)+\nonumber \\
 &&\sum_{k=1}^{N-1} (b^\prime_{k}  { \mbox{Tr}}(\gamma_{2k})+
 {\tilde b}^\prime_{k}  { \mbox{Tr}}({\tilde \gamma}_{2k}))+\nonumber \\
 && \sum_{k=1}^{N-1} ({\cal K}_k+{\cal K}_{k+N})~.
\end{eqnarray}
Here $b^\prime_k \equiv b_k \epsilon_k +b_{k+N} \epsilon_{k+N}$, and 
similarly for ${\tilde b}^\prime_k$. From the above expressions it is clear 
that in order for the twisted tadpoles to cancel it must be the case that 
they cancel in ${\cal T}_{\mbox{{\small odd}}}$ and 
${\cal T}_{\mbox{{\small even}}}$ independently. In fact, the tadpoles must 
factor into sums of perfect squares. Thus, for 
${\cal T}_{\mbox{{\small odd}}}$ we have
\begin{equation}
 {\cal T}_{\mbox{{\small odd}}} =\sum_{k=1}^{N} a_{2k-1} ( 
 { \mbox{Tr}}(\gamma_{2k-1})-{\tilde a}_{2k-1}  { \mbox{Tr}}
({\tilde \gamma}_{2k-1}))^2~,
\end{equation}
where we have used the identity $a_k{\tilde a}_k=1$ which follows from the 
$T$-duality of $D9$- and $D5$-branes. The $T$-duality also implies that we 
can set (up to an irrelevant overall phase) 
${ \mbox{Tr}}(\gamma_{2k-1})={ \mbox{Tr}}({\tilde \gamma}_{2k-1})$, and as 
an immediate consequence we get the following twisted tadpole cancellation 
conditions:
\begin{equation}
  { \mbox{Tr}}(\gamma_{Rg^k})
={ \mbox{Tr}}({\tilde \gamma}_{Rg^k})=0~,~~~k=0,...,N-1.
\end{equation}
The rest of the twisted tadpole cancellation conditions are for the 
Chan-Paton matrices 
${ \mbox{Tr}}(\gamma_{g^k})={ \mbox{Tr}}({\tilde \gamma}_{g^k})$, 
$k=1,...,N-1$. These tadpole cancellation conditions are exactly the same as 
for the ${\bf Z}_N$ orbifold with $N$ odd, and we have already discussed them 
in the previous section. 

{}It is clear now how to construct the Chan-Paton matrices for 
${\bf Z}_2 \otimes {\bf Z}_N$ orbifolds ($N$ is odd). The Chan-Paton 
matrices for the $D9$- and $D5$-branes are the same. Thus, we will confine 
our attention to the $D9$-brane matrices. Those for the twists $g^k$ are 
given in the previous section. The matrix  
$\gamma_R={\mbox{diag}}(i~(8~{\mbox{times}}),-i~(8~{\mbox{times}}))$. 
Also, $\gamma_{Rg^k}=\gamma_R \gamma_{g^k}$. The form of $ \gamma_{g^k}$ 
is then fixed (up to equivalent representations) by the constraints on 
${ \mbox{Tr}}(\gamma_{g^k})$ (this fixes the diagonal elements up to certain 
permutations) and ${ \mbox{Tr}}(\gamma_{Rg^k})$ (this fixes the positions of 
the diagonal elements in $ \gamma_{g^k}$ in the basis where 
$\gamma_R={\mbox{diag}}(i~(8~{\mbox{times}}),-i~(8~{\mbox{times}}))$; 
there is still a residual permutational symmetry, but the latter is 
irrelevant).

{}Let us illustrate the above conditions by the ${\bf Z}_6$ orbifold 
considered in this paper. The twists read:
\begin{eqnarray}
 T_{g}&=&(1/3,1/3,1/3\vert\vert 1/3,1/3,1/3)~,\\
 T_{R}&=&(1/2,1/2,0\vert\vert 1/2,1/2,0)~.
\end{eqnarray}
We have the following Chan-Paton matrices:
\begin{eqnarray}
 {\gamma_R}&=&{\mbox{diag}}(i~(8~{\mbox{times}}),-i~(8~{\mbox{times}}))~,\\
 {\gamma}_{g}&=&{\mbox{diag}}(\exp(2\pi i/3)~(3~{\mbox{times}}),
 \exp(-2\pi i/3)~(3~{\mbox{times}}), 1~(2~{\mbox{times}}),\nonumber\\
 &&~~~\exp(2\pi i/3)~(3~{\mbox{times}}),
 \exp(-2\pi i/3)~(3~{\mbox{times}}), 1~(2~{\mbox{times}}))~,\\
 {\gamma}_{Rg}&=&{\mbox{diag}}(\exp(-\pi i/6)~(3~{\mbox{times}}),
 \exp(-5\pi i/6)~(3~{\mbox{times}}), i~(2~{\mbox{times}}),\nonumber\\
 &&~~~\exp(5\pi i/6)~(3~{\mbox{times}}),
 \exp(\pi i/6)~(3~{\mbox{times}}), -i~(2~{\mbox{times}}))~. 
\end{eqnarray}
Thus, the gauge group is $U(6)\otimes U(6)\otimes U(4)$ in the 99 sector. 
It is the same in the 55 sector if all the $D5$-branes are located at the 
same fixed point.

{}Note that although so far we have discussed only the case of 
${\bf Z}_2 \otimes {\bf Z}_N$ orbifolds ($N$ is odd), the tadplole 
cancellation conditions are staightforward to generalize to 
${\bf Z}_2 \otimes {\bf Z}_{n_1} \otimes {\bf Z}_{n_2} \otimes \cdots 
\otimes {\bf Z}_{n_k}$ orbifolds, where $n_1,n_2,...,n_k$ are odd integers. 
(Here we imply type I orbifolds with $D9$- and $D5$-branes only.) It is also 
straightforward to consider cases where instead of ${\bf Z}_2$ we have 
${\bf Z}_{2^m}$, such as ${\bf Z}_4$, ${\bf Z}_8$ and ${\bf Z}_{12}$.

%%%%%%%%%%%%%Table I %%%%%%%%
%%%%%%%%%%%%%%%%%%%%%%%%%%%%%%%%%%%%%%%%%%%%%%%%%%%%%%%%%%%%%%%%%%%%%%%%%%%%%%%
\begin{table}[t]
\begin{tabular}{|c|c|l|l|}
%%%%%%%%%%%%%%%%%%%%%%%%%%%%%%%%%%%%%%%%%%%%%%%%%%%%%%%%%%%%%%%%%%%%%%%%%%%%
Sector & Field & $SU(4)\otimes SU(4)\otimes SU(4)\otimes SO(8)\otimes U(1)^3$ 
       & Comments \\
\hline
%%%%%%%%%%%%%%%%%%%%%%%%%%%%%%%%%%%%%%%%%%%%%%%%%%%%%%%%%%%%%%%%%%%%%%%%%%%%
Closed & & &\\
Untwisted & $\phi_{a}$ & $3({\bf 1}, {\bf 1}, {\bf 1}, {\bf 1})(0,0,0)_L$ 
          & $a=1,2,3$\\
\hline
%%%%%%%%%%%%%%%%%%%%%%%%%%%%%%%%%%%%%%%%%%%%%%%%%%%%%%%%%%%%%%%%%%%%%%%%%%%%
       & $S^{1k}_{\alpha \beta}$ 
       & $18({\bf 1}, {\bf 1}, {\bf 1}, {\bf 1})(0,0,0)_L$ 
         & $k=1,2$, $\alpha,\beta=1$ to $3$ \\
Closed & $S^{2k}_{\beta \gamma}$ 
       & $18({\bf 1}, {\bf 1}, {\bf 1}, {\bf 1})(0,0,0)_L$ 
         & $k=1,2$, $\beta, \gamma=1$ to $3$ \\
Twisted       & $S^{3k}_{\alpha \gamma}$ 
       & $18({\bf 1}, {\bf 1}, {\bf 1}, {\bf 1})(0,0,0)_L$ 
         & $k=1,2$, $\alpha, \gamma=1$ to $3$ \\
       & $S_{\alpha \beta \gamma}$
       & $27({\bf 1}, {\bf 1}, {\bf 1}, {\bf 1})(0,0,0)_L$
         & $\alpha, \beta, \gamma=1$ to $3$ \\
\hline
%%%%%%%%%%%%%%%%%%%%%%%%%%%%%%%%%%%%%%%%%%%%%%%%%%%%%%%%%%%%%%%%%%%%%%%%%%%%
    & $P_1$ & $(\overline{\bf 4}, {\bf 1}, {\bf 1}, {\bf 8}_v)(-1,0,0)_L$ & \\
    & $P_2$ & $({\bf 1}, {\bf 4}, {\bf 1}, {\bf 8}_v)(0,+1,0)_L$ & \\
    & $P_3$ & $({\bf 1}, {\bf 1}, {\bf 4}, {\bf 8}_v)(0,0,+1)_L$ & \\
&$Q_1$&$({\bf 1}, \overline{\bf 4}, \overline{\bf 4}, {\bf 1})(0,-1,-1)_L$ & \\
Open &$Q_2$&$({\bf 4}, {\bf 1}, \overline{\bf 4}, {\bf 1})(+1,0,-1)_L$ & \\
&$Q_3$&$({\bf 4}, \overline{\bf 4}, {\bf 1}, {\bf 1})(+1,-1,0)_L$ &\\
& $\Phi_1$ & $({\bf 6}, {\bf 1}, {\bf 1}, {\bf 1})(+2,0,0)_L$ & \\
& $\Phi_2$ & $({\bf 1}, {\bf 6}, {\bf 1}, {\bf 1})(0,-2,0)_L$ & \\
& $\Phi_3$ & $({\bf 1}, {\bf 1}, {\bf 6}, {\bf 1})(0,0,-2)_L$ &\\
%%%%%%%%%%%%%%%%%%%%%%%%%%%%%%%%%%%%%%%%%%%%%%%%%%%%%%%%%%%%%%%%%%%%%%%%%%%
\end{tabular}
%%%%%%%%%%%%%%%%%%%%%%%%%%%%%%%%%%%%%%%%%%%%%%%%%%%%%%%%%%%%%%%%%%%%%%%%%%%
\caption{The massless spectrum of the type I model with $N=1$ space-time 
supersymmetry and gauge group 
$SU(4)\otimes SU(4) \otimes SU(4) \otimes SO(8) \otimes U(1)^3$ discussed in 
section II. The gravity, dilaton and gauge supermultiplets are not shown.}  
\end{table}
%%%%%%%%%%%%%%%%%%%%%%%%%%%%%%%%%%%%%%%%%%%%%%%%%%%%%%%%%%%%%%%%%%%%%%%%%%%%%%%

%%%%%%%%%%%%%Table II %%%%%%%%
%%%%%%%%%%%%%%%%%%%%%%%%%%%%%%%%%%%%%%%%%%%%%%%%%%%%%%%%%%%%%%%%%%%%%%%%%%%%%%%
\begin{table}[t]
\begin{tabular}{|c|c|l|l|l|}
%%%%%%%%%%%%%%%%%%%%%%%%%%%%%%%%%%%%%%%%%%%%%%%%%%%%%%%%%%%%%%%%%%%%%%%%%%%%
Sector & Field 
& $SU(4)^{3} \otimes SO(8) \otimes U(1)^3$ 
& $(H_1,H_2,H_3)_{-1}$ & $(H_1,H_2,H_3)_{-1/2}$ \\
\hline
%%%%%%%%%%%%%%%%%%%%%%%%%%%%%%%%%%%%%%%%%%%%%%%%%%%%%%%%%%%%%%%%%%%%%%%%%%%%
  & $\phi_{1}$ & $({\bf 1}, {\bf 1}, {\bf 1}, {\bf 1})(0,0,0)_L$ 
  & $(0,+1,0)$  & $(+{1\over 2},+{1\over 2},+{1\over 2})$ \\
  & $\phi_{2}$ & $({\bf 1}, {\bf 1}, {\bf 1}, {\bf 1})(0,0,0)_L$ 
  & $(-1,0,0)$  & $(-{1\over 2},-{1\over 2},+{1\over 2})$ \\
  & $\phi_{3}$ & $({\bf 1}, {\bf 1}, {\bf 1}, {\bf 1})(0,0,0)_L$ 
  & $(0,0,-1)$  & $(+{1\over 2},-{1\over 2},-{1\over 2})$   \\
    & $P_1$ & $(\overline{\bf 4}, {\bf 1}, {\bf 1}, {\bf 8}_v)(-1,0,0)_L$ 
    & $(0,+1,0)$  & $(+{1\over 2},+{1\over 2},+{1\over 2})$ \\
    & $P_2$ & $({\bf 1}, {\bf 4}, {\bf 1}, {\bf 8}_v)(0,+1,0)_L$  
    & $(-1,0,0)$  & $(-{1\over 2},-{1\over 2},+{1\over 2})$ \\
Untwisted    
    & $P_3$ & $({\bf 1}, {\bf 1}, {\bf 4}, {\bf 8}_v)(0,0,+1)_L$  
    & $(0,0,-1)$  & $(+{1\over 2},-{1\over 2},-{1\over 2})$   \\
&$Q_1$&$({\bf 1}, \overline{\bf 4}, \overline{\bf 4}, {\bf 1})(0,-1,-1)_L$
      & $(0,+1,0)$  & $(+{1\over 2},+{1\over 2},+{1\over 2})$ \\
&$Q_2$&$({\bf 4}, {\bf 1}, \overline{\bf 4}, {\bf 1})(+1,0,-1)_L$
& $(-1,0,0)$  & $(-{1\over 2},-{1\over 2},+{1\over 2})$ \\
&$Q_3$&$({\bf 4}, \overline{\bf 4}, {\bf 1}, {\bf 1})(+1,-1,0)_L$
& $(0,0,-1)$  & $(+{1\over 2},-{1\over 2},-{1\over 2})$   \\
& $\Phi_1$ & $({\bf 6}, {\bf 1}, {\bf 1}, {\bf 1})(+2,0,0)_L$
& $(0,+1,0)$  & $(+{1\over 2},+{1\over 2},+{1\over 2})$ \\
& $\Phi_2$ & $({\bf 1}, {\bf 6}, {\bf 1}, {\bf 1})(0,-2,0)_L$ 
& $(-1,0,0)$  & $(-{1\over 2},-{1\over 2},+{1\over 2})$ \\
& $\Phi_3$ & $({\bf 1}, {\bf 1}, {\bf 6}, {\bf 1})(0,0,-2)_L$
 & $(0,0,-1)$  & $(+{1\over 2},-{1\over 2},-{1\over 2})$   \\
\hline
%%%%%%%%%%%%%%%%%%%%%%%%%%%%%%%%%%%%%%%%%%%%%%%%%%%%%%%%%%%%%%%%%%%%%%%%%%%%
      & $S_{\alpha \beta}^{1 +}$ 
      & $9({\bf 1}, {\bf 1}, {\bf 1}, {\bf 1})(+4/3, +4/3, 0)_L$ 
& $(-{1\over 3},+{2\over 3},0)$ 
& $(+{1\over 6},+{1\over 6},+{1\over 2})$ \\
Twisted   & $S_{\alpha \beta}^{1 -}$ 
      & $9({\bf 1}, {\bf 1}, {\bf 1}, {\bf 1})(-4/3, -4/3, 0)_L$ 
& $(-{2\over 3},+{1\over 3},0)$ 
& $(-{1\over 6},-{1\over 6},+{1\over 2})$ \\
$T_3$, $2 T_3$
   & $T_{\alpha \beta}^{1+}$ 
   & $9({\bf 1}, {\bf 6}, {\bf 1}, {\bf 1})(+4/3, -2/3, 0)_L$ 
& $(-{1\over 3},+{2\over 3},0)$ 
& $(+{1\over 6},+{1\over 6},+{1\over 2})$ \\
   & $T_{\alpha \beta}^{1-}$ 
   & $9({\bf 6}, {\bf 1}, {\bf 1}, {\bf 1})(+2/3, -4/3, 0)_L$ 
& $(-{2\over 3},+{1\over 3},0)$ 
& $(-{1\over 6},-{1\over 6},+{1\over 2})$ \\
\hline
%%%%%%%%%%%%%%%%%%%%%%%%%%%%%%%%%%%%%%%%%%%%%%%%%%%%%%%%%%%%%%%%%%%%%%%%%%%%
      & $S_{\beta \gamma}^{2 +}$ 
      & $9({\bf 1}, {\bf 1}, {\bf 1}, {\bf 1})(+4/3, 0, +4/3)_L$ 
& $(0,+{2\over 3},-{1\over 3})$ 
& $(+{1\over 2},+{1\over 6},+{1\over 6})$ \\
Twisted   & $S_{\beta \gamma}^{2 -}$ 
      & $9({\bf 1}, {\bf 1}, {\bf 1}, {\bf 1})(-4/3, 0, -4/3)_L$ 
& $(0,+{1\over 3},-{2\over 3})$ 
& $(+{1\over 2},-{1\over 6},-{1\over 6})$ \\
$T_3^{\prime}$, $2 T_3^{\prime}$
   & $T_{\beta \gamma}^{2 +}$ 
   & $9({\bf 1}, {\bf 1}, {\bf 6}, {\bf 1})(+4/3, 0, -2/3)_L$ 
& $(0,+{2\over 3},-{1\over 3})$ 
& $(+{1\over 2},+{1\over 6},+{1\over 6})$ \\
   & $T_{\beta \gamma}^{2-}$ 
   & $9({\bf 6}, {\bf 1}, {\bf 1}, {\bf 1})(2/3, 0, -4/3)_L$ 
& $(0,+{1\over 3},-{2\over 3})$ 
& $(+{1\over 2},-{1\over 6},-{1\over 6})$ \\
\hline
%%%%%%%%%%%%%%%%%%%%%%%%%%%%%%%%%%%%%%%%%%%%%%%%%%%%%%%%%%%%%%%%%%%%%%%%%%%
      & $S_{\alpha \gamma}^{3 +}$ 
      & $9({\bf 1}, {\bf 1}, {\bf 1}, {\bf 1})(0,+4/3,-4/3)_L$ 
& $(-{1\over 3},0,-{2\over 3})$ 
& $(+{1\over 6},-{1\over 2},-{1\over 6})$ \\
Twisted   & $S_{\alpha \gamma}^{3 -}$ 
      & $9({\bf 1}, {\bf 1}, {\bf 1}, {\bf 1})(0,-4/3,+4/3)_L$ 
& $(-{2\over 3},0,-{1\over 3})$ 
& $(-{1\over 6},-{1\over 2},+{1\over 6})$ \\
$T_3+2 T_3^{\prime}$, $2 T_3 + T_3^{\prime}$
   & $T_{\alpha \gamma}^{3 +}$ 
   & $9({\bf 1}, {\bf 6}, {\bf 1}, {\bf 1})(0, -2/3, -4/3)_L$ 
& $(-{1\over 3},0,-{2\over 3})$ 
& $(+{1\over 6},-{1\over 2},-{1\over 6})$ \\
   & $T_{\alpha \gamma}^{3-}$ 
   & $9({\bf 1}, {\bf 1}, {\bf 6}, {\bf 1})(0, -4/3, -2/3)_L$ 
& $(-{2\over 3},0,-{1\over 3})$ 
& $(-{1\over 6},-{1\over 2},+{1\over 6})$ \\
\hline
%%%%%%%%%%%%%%%%%%%%%%%%%%%%%%%%%%%%%%%%%%%%%%%%%%%%%%%%%%%%%%%%%%%%%%%%%%%
Twisted & $S_{\alpha \beta \gamma}$ 
        & $27({\bf 1}, {\bf 1}, {\bf 1}, {\bf 1})(-4/3,+4/3,+4/3)_L$ 
& $(-{1\over 3},+{1\over 3},-{1\over 3})$ 
& $(+{1\over 6},-{1\over 6},+{1\over 6})$ \\
$T_3+T_3^{\prime}$, $2 T_3 + 2 T_3^{\prime}$
   & $T_{\alpha \beta \gamma}$ 
   & $27({\bf 1}, {\bf 1}, {\bf 1}, {\bf 8_s})(2/3, -2/3, -2/3)_L$ 
& $(-{1\over 3},+{1\over 3},-{1\over 3})$ 
& $(+{1\over 6},-{1\over 6},+{1\over 6})$ \\
%%%%%%%%%%%%%%%%%%%%%%%%%%%%%%%%%%%%%%%%%%%%%%%%%%%%%%%%%%%%%%%%%%%%%%%%%%%
\end{tabular}
%%%%%%%%%%%%%%%%%%%%%%%%%%%%%%%%%%%%%%%%%%%%%%%%%%%%%%%%%%%%%%%%%%%%%%%%%%%
\caption{The massless spectrum of the heterotic model with $N=1$ space-time 
supersymmetry and gauge group 
$SU(4)\otimes SU(4) \otimes SU(4) \otimes SO(8) \otimes U(1)^3$ 
discussed in section III. The $H$-charges in both the $-1$ picture and
the $-1/2$ picture are also given. The gravity, dilaton and gauge 
supermultiplets are 
not shown.}  
\end{table}
%%%%%%%%%%%%%%%%%%%%%%%%%%%%%%%%%%%%%%%%%%%%%%%%%%%%%%%%%%%%%%%%%%%%%%%%%%%%%%%

%%%%%%%%%%%%%Table III %%%%%%%%
%%%%%%%%%%%%%%%%%%%%%%%%%%%%%%%%%%%%%%%%%%%%%%%%%%%%%%%%%%%%%%%%%%%%%%%%%%%%%%%
\begin{table}[t]
\begin{tabular}{|c|l|l|l|}
%%%%%%%%%%%%%%%%%%%%%%%%%%%%%%%%%%%%%%%%%%%%%%%%%%%%%%%%%%%%%%%%%%%%%%%%%%%%
 Sector & $[SU(6)\otimes SU(6)\otimes SU(4)\otimes U(1)^3]^2$
        & $(H_1,H_2,H_3)_{-1}$ & $(H_1,H_2,H_3)_{-1/2}$ \\
\hline
%%%%%%%%%%%%%%%%%%%%%%%%%%%%%%%%%%%%%%%%%%%%%%%%%%%%%%%%%%%%%%%%%%%%%%%%%%%%
Closed & & &\\
Untwisted & $5({\bf 1}, {\bf 1}, {\bf 1}; {\bf 1}, {\bf 1}, {\bf 1})
(0,0,0;0,0,0)_L$  & & \\
\hline
%%%%%%%%%%%%%%%%%%%%%%%%%%%%%%%%%%%%%%%%%%%%%%%%%%%%%%%%%%%%%%%%%%%%%%%%%%%%
Closed &  & & \\
${\bf Z}_3$ Twisted  & $15({\bf 1}, {\bf 1}, {\bf 1}; {\bf 1}, {\bf 1}, 
{\bf 1})
(0,0,0;0,0,0)_L$ & & \\
\hline
Closed & & &  \\
${\bf Z}_6$ Twisted  & $3({\bf 1}, {\bf 1}, {\bf 1}; {\bf 1}, {\bf 1}, {\bf 1})
(0,0,0;0,0,0)_L$ & & \\
\hline
Closed & & &  \\
${\bf Z}_2$ Twisted  & $11({\bf 1}, {\bf 1}, {\bf 1}; {\bf 1}, {\bf 1}, 
{\bf 1})
(0,0,0;0,0,0)_L$ & & \\
\hline
%%%%%%%%%%%%%%%%%%%%%%%%%%%%%%%%%%%%%%%%%%%%%%%%%%%%%%%%%%%%%%%%%%%%%%%%%%%%
           & $2({\bf 15},{\bf 1},{\bf 1};{\bf 1},{\bf 1},{\bf 1})
(+2,0,0;0,0,0)_L$ & $(+1,0,0)$ & $(+{1\over 2},-{1\over 2},-{1\over 2})$ \\
           &      & $(0,+1,0)$ & $(-{1\over 2},+{1\over 2},-{1\over 2})$ \\   
           & $2({\bf 1},\overline{\bf 15}, {\bf 1}; {\bf 1},{\bf 1},{\bf 1})
(0,-2,0;0,0,0)_L$ & $(+1,0,0)$ & $(+{1\over 2},-{1\over 2},-{1\over 2})$ \\
           &      & $(0,+1,0)$ & $(-{1\over 2},+{1\over 2},-{1\over 2})$ \\ 
           & $2(\overline{\bf 6},{\bf 1},\overline{\bf 4};{\bf 1},{\bf 1},
{\bf 1})(-1,0,-1;0,0,0)_L$ & $(+1,0,0)$ 
& $(+{1\over 2},-{1\over 2},-{1\over 2})$ \\
           &      & $(0,+1,0)$ & $(-{1\over 2},+{1\over 2},-{1\over 2})$ \\ 
Open $99$ & $2({\bf 1},{\bf 6},{\bf 4};{\bf 1},{\bf 1},{\bf 1})
(0,+1,+1;0,0,0)_L$ & $(+1,0,0)$ & $(+{1\over 2},-{1\over 2},-{1\over 2})$ \\
           &       & $(0,+1,0)$ & $(-{1\over 2},+{1\over 2},-{1\over 2})$ \\ 
           & $({\bf 6},\overline{\bf 6},{\bf 1};{\bf 1},{\bf 1},{\bf 1})
(+1,-1,0;0,0,0)_L$ & $(0,0,+1)$ & $(-{1\over 2},-{1\over 2},+{1\over 2})$ \\
           & $(\overline{\bf 6},{\bf 1},{\bf 4};{\bf 1},{\bf 1},{\bf 1})
(-1,0,+1;0,0,0)_L$ & $(0,0,+1)$ & $(-{1\over 2},-{1\over 2},+{1\over 2})$ \\
           & $({\bf 1},{\bf 6},\overline{\bf 4};{\bf 1},{\bf 1},{\bf 1})
(0,+1,-1;0,0,0)_L$ & $(0,0,+1)$ & $(-{1\over 2},-{1\over 2},+{1\over 2})$ \\ 
\hline
%%%%%%%%%%%%%%%%%%%%%%%%%%%%%%%%%%%%%%%%%%%%%%%%%%%%%%%%%%%%%%%%%%%%%%%%%%%
           & $2({\bf 1},{\bf 1},{\bf 1};{\bf 15},{\bf 1},{\bf 1})
(0,0,0;+2,0,0)_L$ & $(+1,0,0)$ & $(+{1\over 2},-{1\over 2},-{1\over 2})$ \\
           &      & $(0,+1,0)$ & $(-{1\over 2},+{1\over 2},-{1\over 2})$ \\ 
           & $2({\bf 1},{\bf 1},{\bf 1};{\bf 1},\overline{\bf 15},{\bf 1})
(0,0,0;0,-2,0)_L$ & $(+1,0,0)$ & $(+{1\over 2},-{1\over 2},-{1\over 2})$ \\
           &      & $(0,+1,0)$ & $(-{1\over 2},+{1\over 2},-{1\over 2})$ \\ 
          & $2({\bf 1},{\bf 1},{\bf 1};\overline{\bf 6},{\bf 1},
\overline{\bf 4})(0,0,0;-1,0,-1)_L$ & $(+1,0,0)$ 
                  & $(+{1\over 2},-{1\over 2},-{1\over 2})$ \\
           &      & $(0,+1,0)$ & $(-{1\over 2},+{1\over 2},-{1\over 2})$ \\
Open  $55$ & $2({\bf 1},{\bf 1},{\bf 1};{\bf 1},{\bf 6},{\bf 4})
(0,0,0;0,+1,+1)_L$ & $(+1,0,0)$ & $(+{1\over 2},-{1\over 2},-{1\over 2})$ \\
           &      & $(0,+1,0)$ & $(-{1\over 2},+{1\over 2},-{1\over 2})$ \\ 
           & $({\bf 1},{\bf 1},{\bf 1};{\bf 6},\overline{\bf 6},{\bf 1})
(0,0,0;+1,-1,0)_L$ & $(0,0,+1)$ & $(-{1\over 2},-{1\over 2},+{1\over 2})$ \\
           & $({\bf 1},{\bf 1},{\bf 1};\overline{\bf 6},{\bf 1},{\bf 4})
(0,0,0;-1,0,+1)_L$ & $(0,0,+1)$ & $(-{1\over 2},-{1\over 2},+{1\over 2})$ \\
           & $({\bf 1},{\bf 1},{\bf 1};{\bf 1},{\bf 6},\overline{\bf 4})
(0,0,0;0,+1,-1)_L$ & $(0,0,+1)$ & $(-{1\over 2},-{1\over 2},+{1\over 2})$ \\
\hline
%%%%%%%%%%%%%%%%%%%%%%%%%%%%%%%%%%%%%%%%%%%%%%%%%%%%%%%%%%%%%%%%%%%%%%%%%%%
           & $({\bf 6},{\bf 1},{\bf 1};{\bf 6},{\bf 1},{\bf 1})
(+1,0,0;+1,0,0)_L$ & $(+{1\over 2},+{1\over 2},0)$ & $(0,0,-{1\over 2})$ \\
           & $({\bf 1},{\bf 6},{\bf 1};{\bf 1},{\bf 1},{\bf 4})
(0,+1,0;0,0,+1)_L$ & $(+{1\over 2},+{1\over 2},0)$ & $(0,0,-{1\over 2})$ \\
Open  $59$ & $({\bf 1},{\bf 1},{\bf 4};{\bf 1},{\bf 6},{\bf 1})
(0,0,+1;0,+1,0)_L$ & $(+{1\over 2},+{1\over 2},0)$ & $(0,0,-{1\over 2})$ \\
           & $(\overline{\bf 6},{\bf 1},{\bf 1};{\bf 1},{\bf 1},
\overline{\bf 4})(-1,0,0;0,0,-1)_L$ 
& $(+{1\over 2},+{1\over 2},0)$ & $(0,0,-{1\over 2})$ \\ 
          & $({\bf 1},\overline{\bf 6},{\bf 1};{\bf 1},\overline{\bf 6},
{\bf 1})(0,-1,0;0,-1,0)_L$ 
& $(+{1\over 2},+{1\over 2},0)$ & $(0,0,-{1\over 2})$ \\ 
          & $({\bf 1},{\bf 1},\overline{\bf 4};\overline{\bf 6},{\bf 1},
{\bf 1})(0,0,-1;-1,0,0)_L$
& $(+{1\over 2},+{1\over 2},0)$ & $(0,0,-{1\over 2})$ \\ 
%%%%%%%%%%%%%%%%%%%%%%%%%%%%%%%%%%%%%%%%%%%%%%%%%%%%%%%%%%%%%%%%%%%%%%%%%%%
\end{tabular}
%%%%%%%%%%%%%%%%%%%%%%%%%%%%%%%%%%%%%%%%%%%%%%%%%%%%%%%%%%%%%%%%%%%%%%%%%%%
\caption{The massless spectrum of the type I ${\bf Z}_6$ orbifold model 
with $N=1$ space-time 
supersymmetry and gauge group 
$[SU(6)\otimes SU(6) \otimes SU(4) \otimes U(1)^3]^2 $
discussed in 
section VI. 
The $H$-charges in both the $-1$ picture and the $-1/2$ picture for states
in the open
string sector are also given. The gravity, dilaton and gauge supermultiplets 
are not shown.}  
\end{table}
%%%%%%%%%%%%%%%%%%%%%%%%%%%%%%%%%%%%%%%%%%%%%%%%%%%%%%%%%%%%%%%%%%%%%%%%%%%%%%%

%%%%%%%%%%%%%%%%%%%%%%%%%%%%%%%%%%%%%%%%%%%%%%%%
\newpage
\begin{figure}[t]
%\hspace*{}
%\vspace*{}
\epsfxsize=16 cm
\epsfbox{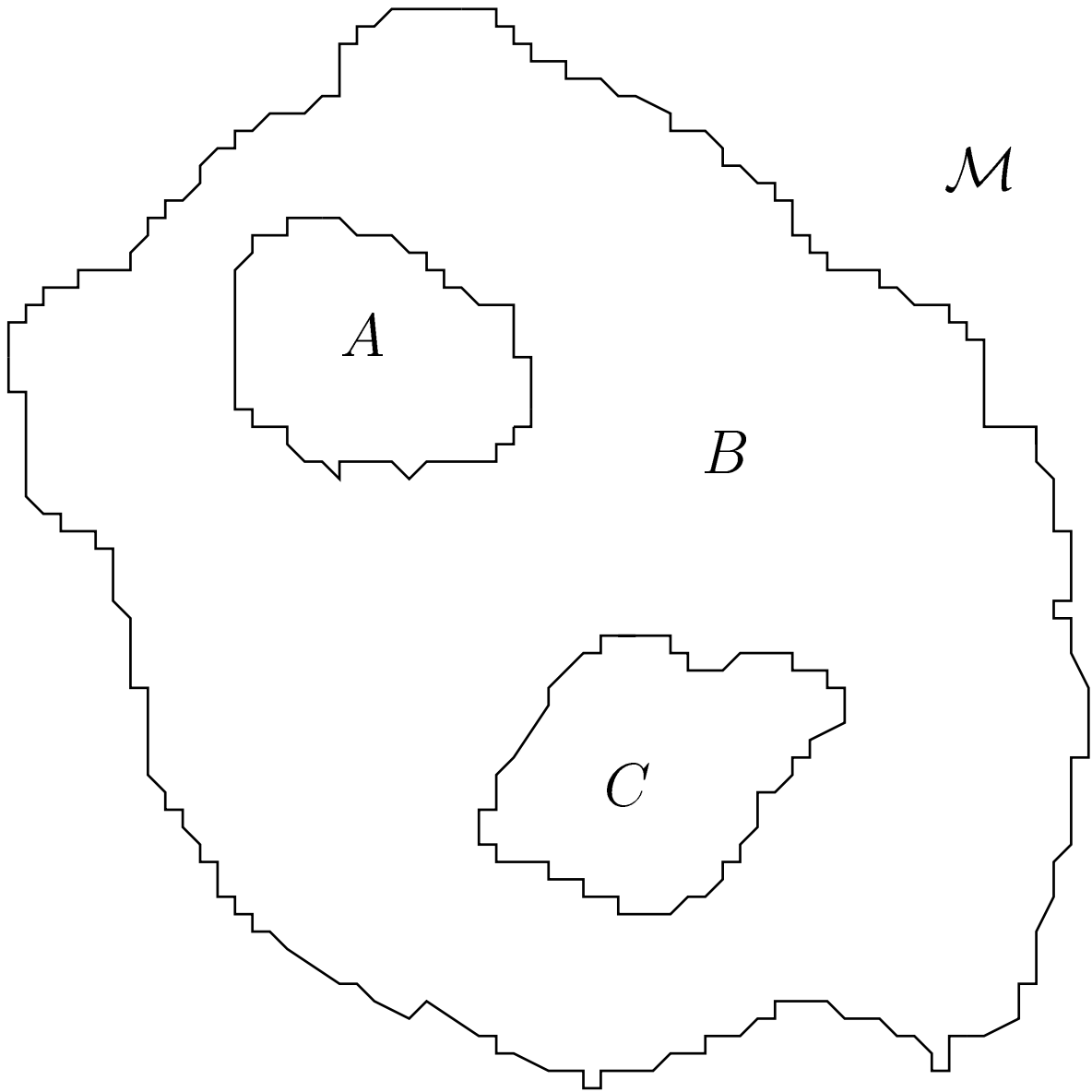}
\caption{A schematic picture of the (perturbative) moduli space ${\cal M}$
of the heterotic ${\bf Z}_3 \otimes {\bf Z}_3$ orbifold. 
This figure is taken from 
Ref [2] where the ${\bf Z}_3$ orbifold is discussed since  
the schematic picture of the moduli space for both 
${\bf Z}_3$ and ${\bf Z}_3 \otimes {\bf Z}_3$ orbifolds are the same. 
Region $A$ is the subspace corresponding to the 
type I model. Region $C$ is the subspace where some or all of the 
$S_{\alpha\beta\gamma}$ vevs are zero and some or all of the 
$T_{\alpha\beta\gamma}$ and $T^{a\pm}_{\alpha\beta}$ fields are massless. 
Region $B$ complements $A$ and $C$ in ${\cal M}$.}
\end{figure}
%%%%%%%%%%%%%%%%%%%%%%%%%%%%%%%%%%%%%%%%%%%%%%%%

\end{document}